\documentstyle[12pt]{article}

\begin{document}


\begin{center}
\Large{\bf Singularity: Raychaudhuri Equation once again}
\end{center}

\begin{center}
Naresh Dadhich\\  
Inter-University Centre for Astronomy \& Astrophysics,
Post Bag 4, Pune~411~007, India \\ 
E-mail: nkd@iucaa.ernet.in
\end{center}

\bigskip

\abstract{I first recount Raychaudhuri's deep involvement with the
singularity problem in general relativity. I then argue that precisely
the same situation has arisen today in loop quantum cosmology as
obtained when Raychaudhuri discovered his celebrated equation. We thus
need a new analogue of the Raychaudhuri equation in quantum gravity.} 
\vskip2pc 

\vskip2pc


\section{Singularity and AKR}

It would not be far from the truth to say that A. K. Raychaudhuri
(AKR) had a fascinatingly engaging love affair with the notion of a
spacetime singularity at the two ends of his research career. One of
his early concerns was to construct a model of a collapsing
homogeneous dust ball (he was unaware of Oppenheimer-Snyder collapse)
and show that nothing prevented the ball from collapsing down to the
centre $r = 0$ and thereby demystify the so called Schwarzschild
singularity at $r = 2M$ ~\cite{akr}. Then he addressed the most
pertinent question of his time: is the cosmic singularity predicted by
the FRW model an artifact of the homogeneity and isotropy of space or
not? As he explained in his reminiscences ~\cite{akr1}, inspired by
the famous G{$\ddot{o}$}del solution, he was looking for a rotating
non-singular solution without closed timelike lines. In the process,
he discovered his celebrated equation ~\cite{ray1} which made the
singularity analysis free of these restrictions. That ultimately led
to the powerful Hawking-Pensrose singularity theorems ~\cite{hp} which
established in a very general setting the inevitability of the
occurrence of singularities in Einstein gravity under reasonable
energy and causality conditions. \\

The other end phase began in the mid 1990s. The singularity theorems
reigned supreme, particularly since the observation of CMBR \cite{pw}
had pointed to a singular birth of the Universe in a big bang. Nothing
could be happier and more persuasive than observation verifying the
prediction of theory. This gave rise to a general belief that
singularities were inevitable in general relativity (GR) so long as
the dynamics were governed by Einstein's equations and moreover
positive energy and causality conditions were respected. However, this
belief was shaken by Senovilla's discovery in 1990 of a singularity
free cosmological solution \cite{seno} which did not violate the
energy and causality conditions. How could such a thing happen? It
brought forth the main suspect in the proofs of the singularity
theorems. Apart from the self-evident assumptions, the theorems also
required the existence of a closed trapped surface. This last
requirement is certainly not so obvious and self-evident as the other
asumptions. That gravity should become so strong in some bound region
of space that even light could not escape from it is a very limiting
assumption. Indeed, where gravity should become how strong ought to be
determined by the field equations rather than by prescription. The
said assumption may, however, be reasonable and justifiable for the
case of the gravitational collapse of an isolated body. We know from
the study of stellar structure that a sufficiently massive body could,
after the exhaustion of its nuclear fuel, ultimately undergo
indefinite collapse and thereby reach the trapped surface limit. In
the case of big bang cosmology what is required is not a trapped
surface but instead sufficient amount of matter distribution for the
focusing of non-spacelike trajectories at a finite proper time in the
past. It is a different matter that the amount required to thermalise
the cosmic background radiation is indeed sufficient for the
convergence of trajectories in the past ~\cite{hp}. Though this
limitation in terms of either a trapped surface or a sufficient amount
of energy density was known to experts in the field, it was not talked
about much, perhaps in the belief that a singularity free solution
would never be found. \\

In the early 1990s, L. K. Patel, Ramesh Tikekar and myself obtained
some singularity free cosmological solutions ~\cite{ptd}, in
particular one with a stiff fluid equation of state $\rho = p$. During
this period, I had several discussions with AKR. We both shared the
view that the assumption of the existence of a closed trapped surface
almost amounted to putting in a singularity. Nothing could come out of
a closed trapped surface nor could the collapse be halted or reversed
inside it without violating energy and causality conditions. Thus a
singularity would become inevitable. He was also not happy with the
genericity condition. In his view it was too complicated and
physically not very illuminating. In the ICGC meeting at IUCAA in
December 1995, Jose Senovilla and I had discussions with him and he
then came out with an insightful comment. He opined that the vanishing
of the space average of physical and kinamatic parameters was required
for singularity free solutions. We were both struck by this comment
which showed a new direction. (See Senovilla's acount in this
volume~\cite{seno1}). \\

AKR had started thinking about singularity free cosmological
solutions, but was not yet quite taken up by them. In November 1996,
there was an IUCAA sponsored workshop on Inhomogemeous Cosmological
Models at North Bengal University, Siliguri. I spoke there on
singularity free cosmological solutions. During the talk, AKR asked
several probing questions and we had a very lively and engaging
discussion. It was indicative of his thought process in trying to
understand and resolve intricate and involved conceptual and physical
issues. It took him a couple of years before he could start working on
the question of the avoidance of the cosmic singularity. This showed
his work ethic - deep and long period of contemplation and thought
before taking up a problem. This venture took him once again to the
question of singularity theorems. He argued that the existence of
singularity free cosmological solutions should be recognized and
proposed the vanishing of the spacetime averages of all the scalars
appearing in the Raychaudhuri equation as a necessary condition for
their existence ~\cite{ray2,ray3,ray4}. He later proved a new
singularity theorem in which he replaced the occurrence of a closed
trapped surface by the non-vanishing of the space averages of all
scalars occurring in the Raychaudhuri equation ~\cite{ray4}. The
vanishing of such space averages was shown to be the key to
singularity free cosmological models. \\

The last paper that AKR wrote was in 2004. In this he attempted to
deduce the Ruiz-Senovilla family ~\cite {rs} of non-singular solutions
for a non-rotating perfect fluid from very general considerations. His
procedure was novel, though not mathematically
rigorous ~\cite{ray5}. A very large family of singularity free
cosmological models including some counter examples to his paper
~\cite{ray5} had also been found ~\cite{jg}. It is however known that
for an imperfect fluid it is easy to construct non-singular and even
oscillating models ~\cite{nd,dr}. The real challenge is, in fact, in
obtaining rotating perfect fluid solutions. Apart from increasing the
mathematical complexity, rotation brings in the question of the
occurrence of closed timelike lines and consequently causality
violation. We have the well-known rotating G{$\ddot{o}$}del universe
which has closed timelike lines. Recall that it was precisely the
G{$\ddot{o}$}del solution which had set him on the singularity
trail. Right at the beginning in the early 1950s, his main aim was to
find a rotating fluid solution, hopefully free of any singularity as
well as of any closed timelike line. Instead, he discovered his
equation. The question remained open and unsolved, however. In fact,
AKR returned once again to it at the end. It is undobtedly one of the
most challenging open problems in classical gravity today. Ironically,
he began and also breathed his last with it on 18 June 2005. \\
 
AKR had the profound insight to have identified the key feature of
non-singular solutions, namely the vanishing of space averages of
physical paramneters. It was the interaction between him and Jose
Senovilla which led to the formulation of this conjecture, though each
of them had a different perspective on it. AKR's attempt to prove it
~\cite{ray4} was not entirely satisfactory and Senovilla has now
proved it ultimately ~\cite{seno1}. This result should rightly be
called the Raychaudhuri-Senovilla theorem. It was perhaps the
limitation of mathematical and analytical tools that AKR had at his
command which came in the way of his proving the theorem
rigorously. Yet, had it not been for his insight, the theorem might
not have been formulated. Therefore it is to AKR's credit that he
showd the right path in understanding singularity free cosmological
solutions. If he had the benefit of the right kind of mathematical
backup in the mid 1950s, he could conceivably have arrived at the
famous singularity theorems. Once again, I believe that the limiting
factor was mathematical technology. With utmost reverence and
affection, I would like to acknowledge this fact in the true spirit of
AKR which embodied academic and itellectual honesty and
objectivity. \\

After recording the story of my understanding and perception of the
AKR-singularity saga, let me change gears in the next section to argue
that his equation in a new avatar is once again badly needed.

\section{Equation once again}

A singularity marks the limiting point of a physical theory. It is
enigmatic and calls for a new theory. In Einstein's GR, gravity is
nothing but the curvature of spacetime. A gravitational singularity
thus means the breakdown of spacetime structure itself and hence the
end of everything.\\

The forces of Newtonian gravity as well as of Maxwell's electric field
diverge and are singular at the central location of the mass/charge
point. This singularity does not disturb the spacetime
background. Rather, it indicates the limit of validity of the
theory. For the electric field, we go over to quantum electrodynamics
to overcome the classical singularity. For gravity, apart from
addressing the singularity, we also need a new theory for the more
basic requirement of making it fully universal. So we have Einstein's
theory of gravitation, namely GR. But in this new theory, the
Newtonian singularity not only persists but in fact attains a more
profound all encompassing proportion. One therefore needs a quantum
theory of gravity which has to address the question of singularity in
the spacetime structure itself. \\

GR made two profound predictions, one of the black hole and the other
of the big bang.  Both harboured singularities. In the former case it
is hidden behind an event horizon and hence is inaccessible to an
external observer. Though the Schwarzschild solution was obtained in
1916 immediately after GR was propounded, its full import as the
representation of a static black hole was not realized as late as the
late 1960s.  A star, collapsing under its own gravity, will go on
collapsing indefinitely upon the exhaustion of all its nuclear fuel
and eventually hit the central singularity $r=0$.  The latter would be
encompassed by a black hole event horizon from which nothing could
come out. Penrose in 1969 pronounced that any singularity occurring in
a gravitational collapse will always be covered by a black hole and
this is known as the Cosmic Censorship Conjecture. Oppenheimer and
Snyder considered the collapse of a homogeneous dust cloud. Their
conclusion is that it collapses down to a singularity covered by a
black hole. \\

In 1924, Friedmann obtained a non-static solution to Einstein's
equations representing an expanding model of the universe. In 1929,
Hubble's observation of receding galaxies lent observational strength
to this model. It was indeed a wonderful marriage of theory and
observation. The matter distribution in the universe was assumed to be
homogeneous and isotropic. It predicted that the universe, which was
now expanding, would have had a singular beginning in a hot big bang
when all matter was concentrated within a very small pointlike
region. \\

An important question then arose. Was this singularity an artifact of
the symmetries of matter distribution, to wit homogeneity and
isotropy, or a generic feature of Einstein's gravity? That is when AKR
came on stage and formulated the singularity issue in all its
generality and obtained his celebrated equation in 1953
~\cite{ray1}. The equation brought to the fore the new feature that
shear as well as pressure contribute positively to gravity, while
rotation goes the other way, as expected. Inspired by the Raychaudhuri
equation, Penrose, Hawking and Geroch then proved in the mid 1960s
their powerful singularity theorems ~\cite{hp} under very general
conditions to establish that singularities are inevitable in GR so
long as some reasonable energy and causality conditions are
satisfied. \\

Within the classical framework, just as in Maxwell electrodynamics,
there is no way to avoid a singularity in Einstein's gravity. That is
why one of the main goals of any quantum gravity theory is to address
the singularity question. We do not yet have a full-fledged theory of
quantum gravity. There are two main attempts. One is string theory
which is based on particle physics. The other is the canonical
quantization scheme of loop quantum gravity which is based on GR. We
shall follow developments in the latter since it directly addresses
the singularity issue. \\

The loop quantum gravity (LQG) idea rests on an important breakthrough
achieved by Abhay Ashtekar in 1986. He discovered new variables in
which Einstein's equations take a polynomial form ~\cite{a86}. Since
then he has spearheaded this approach. This effort is, however,
pursued by a comparatively small but highly committed and talented
team of researchers ~\cite{aa}. Even in the absence of a full theory,
it is instructive and insightful to apply this developing theory to
idealized special cases and probe for possible signatures of quantum
gravity effects in astrophysical and cosmological observations. Such
applications, howsoever tentative, serve as good testbeds for the
evolving theory in regard to its right orientation and direction.
With that in view, Martin Bojowald and others have, for the past few
years, been examining cosmological applications of loop quantum
gravity. Such efforts have led to the subject of loop quantum
cosmology (LQC) where one considers the symmetry reduced mini
superspace and then carries out loop quantum calculations for specific
problems of big-bang cosmology, cosmic microwave background radiation
and gravitational collapse ~\cite{mb,aps,ab,gd,gjs,stm}. The most
pertinent question is whether one could have some observational
imprint or signature of quantum gravity effects. \\

Recently, there have been a couple of LQC based calculations to look
for quantum effects in astrophysical and cosmological scenarios.  The
first observable effect of LQC was studied on CMBR in 2003 by Shinji
Tsujikawa, Parampreet Singh and Roy Maartens ~\cite{stm}. It turns out
that quantum gravity effects could avoid the big bang singularity and
there could be a causal passage through it. Its imprint could
therefore be seen on the CMBR spectra. In a recent paper, Abhay
Ashtekar, T Pawlowski and Parampreet Singh further illuminate on the
quantum nature of the big bang ~\cite{aps}. In this context, it is
worth recalling one of the first attempts made by T Padmanabhan and J
V Narlikar in the 1980s towards avoiding the big bang singularity by
quantizing conformal degree of freedom ~\cite{pn}. The pertinent
question is how to justify in the extreme high energy regime highly
restricted degrees of freedom or the reduced mini superspace of
LQC? Both considerations suffer from this lack of justification. It
should however be noted that, though there is no rigorous derivation
of LQC from LQG, the former does have a good theoretical backup with
proper caveats. There has also been a consideration of the collapse of
a homogeneous scalar field by R Goswami, P Joshi and Parampreet Singh
where LQC effects make the central singularity evaporate away as
radiation ~\cite{gjs}. An observational signature of such a quantum
evaporation of a naked singularity may be a pulse of intense radiation
such as a gamma ray burst (GRB). \\

We now come back to the old question: is such an avoidance of the big
bang or of a collapse singularity an artifact of the symmetry reduced
mini superspace or is it generic to LQC?. Near the singularity,
curvatures are divergingly high. So it would not be possible to
truthfully sustain the assumption of a reduced mini superspace. Once
again, we need another AKR today to find a new avatar of his
equation. In other words, a new Raychaudhuri equation is urgently
required in loop quantum gravity. It may, like the old one, show the
way to new general quantum singularity (avoidance) theorems. \\

It is true that LQC deals with highly restrictive and idealized cases.
One should, however, note that such idealized models have the uncanny
knack of picking up the physical essence and innate characteristic of
real life phenomena. There are several such examples in gravitational
theories. The most famous one is, of course, that of the FRW model
predicting a big bang singularity and similarly the Oppenheimer-Snyder
collapse of a homogeneous dust ball. Despite being a highly idealized
case, the model does carry quite truthfully the signature of a general
collapse phenomenon. Similar is the case of the Schwarzschild interior
solution with a uniform density assumed for the interior of a
star. Such an assumption is physically unacceptable since a uniform
density would give rise to an infinite sound speed. Nevertheless, the
model picks up all the essential features of the stellar interior
correctly. It is remarkable that, even when such considerations are
highly restricted, idealized and not even entirely physically
acceptable, they often correctly indicate general as well as generic
features. Such may as well be the case for the highly idealized LQC
toy models indicating the quantum avoidance of the big bang and
collapse singularities. As already emphasized, what is needed is a
Raychaudhuri equation for quantum gravity. It may bring forth some new
features like the development of a negative pressure as the
singularity is approached. Until then we have to make do with
tentative results which may only be indicative of what the full
quantum gravity will ultimately establish. \\

Finally, all this should be most satisfying and pleasing to one man,
Abhay Ashtekar. He made the path breaking discovery two decades back
of his famous new variables which set things on track, leading to Loop
Quantum Gravity. This idea has now matured sufficiently to make
contact with observations. It has indeed been a long and ardous
journey. But, at the end of the day, nothing could please one more
than to see the clicking of what one had set out do. Admittedly, we
are far from a complete theory of quantum gravity and LQG has a long
way to go yet. What is important is that it seems to be on the right
track ~\cite{an}. Recent works of Rovelli, Speziale and others
~\cite{cr,bm,ls} on graviton propagators in LQG and of Smolin et al
~\cite{btl} on the origin of the standard model of particle physics
from the quantum nature of geometry are indeed very exciting.

\section{Acknowledgement} I thank Jose Senovilla for recounting and
reflecting on our discussion with AKR and also thank Parampreet Singh
and Pankaj Joshi for clarifying comments. I also thank my fellow editor of 
this volume, Probir Roy for reading the manuscript and making changes which 
have made it more impactful.




\begin{thebibliography}{}   
\bibitem{akr}A. K. Raychaudhuri, Phys. Rev. {\bf84}, 166 (1951).
\bibitem{akr1} A. K Raychaudhuri, A Little Reminiscence in Singularities, Black Holes and Cosmic Censorship, ed. P. S. Joshi (IUCAA,1996).
\bibitem{ray1} A. K. Raychaudhuri, Phys. Rev. {\bf90}, 1123 (1955).
\bibitem{hp} S. W.Hawking and G. F. R. Ellis, The Large Scale Structure of space-time (Cambridge Univeristy Press, 1973). 
\bibitem{pw} A. A. Penzias and R. W. Wilson, Astrophys. J. {\bf142}, 419 
(1965).
\bibitem{seno} J. M. M. Senovilla, Phys. Rev. Lett. {\bf 64}, 2219 (1990).
\bibitem{ptd} N. Dadhich, L. K. Patel and R Tikekar, Pramana, {\bf 44}, 303 (1995).
\bibitem{seno1} J. M. M. Senovilla, in this volume.
\bibitem{ray2} A. K. Raychaudhuri, Phys. Rev. Lett. {\bf 80}, 654 (1998).
\bibitem{ray3}A. K. Raychaudhuri, A Fresh Look at the Singularity Problem, in 
{\it The Universe}, eds. N. Dadhich and A. K. Kembhavi, (Kluwer, 2000).
\bibitem{ray4} A. K. Raychaudhuri, Mod. Phys. Lett. {\bf A15}, 319 (2000).
\bibitem{rs} E. Ruiz and J. M. M. Senovilla, Phys. Rev. {\bf D45}, 1995 (1992).
\bibitem{ray5} A. K. Raychaudhuri, Gen. Relativ. Grav. {\bf36}, 343 (2004).
\bibitem{jg} L. Fernandez-Jambrina and L. M. Gonzalez-Romero, Phys. Rev 
{\bf D66}, 024027 (2002).  
\bibitem{nd} N. Dadhich, J. Astrophys. Astro. {\bf18}, 343 (1997).
\bibitem{dr} N. Dadhich and A. K. Raychaudhuri, Mod. Phys. Lett. {\bf A14}, 
2135 (1999).
\bibitem{a86} A. Ashtekar, Phys. Rev.Lett. {\bf 57}, 2244 (1986).
\bibitem{aa}  A. Ashtekar, Gravity, Geometry and the Quantum, gr-qc/0605011. 
\bibitem{mb} M. Bojowald, Loop, Quantim Cosmology, Living Rev. Rel. {\bf 8}, 11 (2005), gr-qc/0601085.
\bibitem{aps} A. Ashtekar, T. Pawlowski and P. Singh. Phys. Rev. Lett. {\bf 96}, 141301 (2006); Phys. Rev. {\bf D73}, 124038 (2006); {\bf D74}, 084003 (2006).
\bibitem{ab} A. Ashtekar and M. Bojowald, Class. Qaunt. Grav. {\bf 22}, 3349 (2005); {\bf 23}, 391 (2006).
\bibitem{gd} G. Date and G.M. Hossain, Phys. Rev. Lett. {\bf 94}, 011302 (2005). 
\bibitem{gjs} R. Goswami, P.S. Joshi and P. Singh, Phys. Rev. Lett. {\bf 96}, 031302 (2006).
\bibitem{stm} S. Tsujikawa, P. Singh and R. Maartens, Class. Quant. Grav. {\bf 21}, 5767 (2004).
\bibitem{pn} T. Padmanabhan in Highlights in Gravitation and Cosmology, eds. B. R. Iyer, A. Kembhavi, J. V. Narlikar and C. V. Visheveshwara (Cambridge University Press, 1988), p. 156.  
\bibitem{an} A. Ashtekar, Physics from Geometry, Nature Physics, {\bf 2}, 725 (2006).
\bibitem{cr} C. Rovelli, Phys. Rev. Lett. {\bf 97}, 151301 (2006).
\bibitem{bm} E. Bianchi, L. Modesto, C. Rovelli and S. Speziale, Class. Qaunt. Grav. {\bf 23}, 6989 (2006).
\bibitem{ls} E. Livine and S. Speziale, JHEP, {\bf 0611}, 092 (2006).
\bibitem{btl} S. O. Bilson-Thompson, F. Markopoulou and L. Smolin, Quantum Gravity and the Standard Model, hep-th/0603022. 


\end{thebibliography}
\end{document}